\documentstyle[12pt]{article}
\begin{document}
\def\lb{\label}
\def\PP{\Pi_{p}(N)}
\def\P{\Pi}
\def\t{\theta}
\def\ti{\theta_i}
\def\tj{\theta_j}
\def\tk{\theta_k}
\def\be{\begin{equation}}
\def\ee{\end{equation}}
\def\d{\partial}
\def\a{\alpha}
\def\s{\sigma}
\def\r{\rho}
\def\rij{\rho_{ij}}
\def\rik{\rho_{ik}}
\def\rjk{\rho_{jk}}
\def\rki{\rho_{ki}}
\def\b{\beta}
\def\g{\gamma}
\def\G{\Gamma}
\def\GG{\Gamma_{p}(N)}
\def\gg{{\bf g}}

\hoffset = -1truecm
\voffset = -2truecm

\title{
{\bf Paragrassmann Algebras with Many Variables}
}

\author{
{\bf
A.T. Filippov\thanks{supported in part by Russian Fund for
Fundamental Research, grant number 93--02--3827}}\\
\normalsize I.N.F.N., Sezione di Torino, via P.Giuria 1, I-10125 Torino
{\bf Italy}\\
{\normalsize and}\\
\normalsize Joint Institute for Nuclear Research, 141980 Dubna,
{\bf Russia}\thanks{Permanent Address} \and
{\bf A.B. Kurdikov
\thanks{supported in part by Russian Fund for
Fundamental Research, grant number 93--02--3827}}\\
\normalsize Joint Institute for Nuclear Research, 141980 Dubna,
{\bf Russia} }

\date{}
\newpage

\maketitle

\begin{abstract}
This is a brief review of our recent work attempted at a generalization
of the Grassmann algebra to the  paragrassmann ones.    The main aim is
constructing an algebraic basis for representing   `fractional'
symmetries appearing in $2D$ integrable models and also introduced
earlier as a natural generalization of supersymmetries.   We have
shown that these algebras are naturally related to quantum groups
with $q = {\rm root \;of \; unity}$. By now we have a general construction
of the  paragrassmann calculus  with one variable  and preliminary
results on deriving a natural generalization of the Neveu--Schwarz--Ramond
algebra.  These results were very recently published in two
papers (in collaboration with A.Isaev).

The main emphasis of this report is on  a new general construction of
paragrassmann algebras with any number of variables, N. It is shown
that for the nilpotency indices  $(p + 1) = 3, 4, 6$ the algebras are
almost as simple as the Grassmann algebra  (for which $(p + 1) = 2$).
 A general algorithm  for deriving algebras with
arbitrary p and N is also given.  However,  it is shown that this
algorithm does not exhaust all possible algebras, and the simplest
example of an `exceptional' algebra is presented for $p = 4, N = 4$.
\end{abstract}

\newpage

We start with a brief report
\footnote{This is partly based on the reports presented to the
Smorodinsky Workshop on Group Theory in Physics, Dubna, July 1993}
on our recent results on constructing
paragrassmann algebras with many variables. A detailed motivation
and our previous results may be found in
Refs.\cite{fik}, \cite{fik1}, \cite{fik2} where references to
other related papers can also be found.
As our approach is distinctly different from other attempts in the same
direction we first outline general principles we use in this
construction. Mathematically, our approach is to derive a correct
para-generalization of the classical fermionic algebra
(i=1, \ldots , N):
\be
\lb{w1}
\t_{i}^2  =  0 \; , \;
\d_{i}^2  =  0 \; .
\ee
\be
\lb{w11}
\t_{i}\t_{j}  =  - \t_{j}\t_{i}   \; , \;
\d_{i}\d_{j}  =  - \d_{j}\d_{i}    \; .
\ee
\be
\lb{w12}
\d_{i}\t_{j}  =  - \t_{j}\d_{i}  \; , \;
\d_{i}\t_{i}  =  - \t_{i}\d_{i} + 1 \; .
\ee
For the moment, the prefix {\it para} means that the first equation
is replaced by the so called $p$-nilpotency condition:
$$
\t_{i}^{p+1} = 0 =\d_{i}^{p+1}\;, \; i = 1,\dots ,N\;.
$$
To see what might be called {\em a correct generalization} of other two
equations we have to discuss them in more detail.

Eqs.(\ref{w1}, \ref{w11}) guarantee that any linear combination of
$\t_{i}$ is nilpotent, and the same is true for $\d_{i}$.
Using Eqs.(\ref{w11}, \ref{w12}) one can change the order of the
products of the variables $\t_{i}$ and
derivatives $\d_{i}$. It is convenient to rewrite Eq.(\ref{w12}) as
\be
\lb{w13}
\d_{i}\t_{j}=\delta_{ij} - \t_{j}\d_{i} \; .
\ee
In this form one easily recognizes the commutation relations between
the Fermi creation and annihilation operators (the only difference is
that $\t$ are not supposed to be Hermitian conjugate of $\d$. Having
in mind this analogy, it is natural to define a vacuum vector $|0>$,
$$\d_{i} |0> = 0 \;, $$
and to construct $2^N$ independent states
$$ \t^{k_N}_{N} \ldots \t^{k_1}_{1} |0> = |k_1 \ldots k_N> \;,$$
where $k_i = 0 \; {\rm or} \; 1 \;.$  With the corresponding dual
(conjugate) vacuum, vanishing under the action of $\t_i$ from the right,
one can construct $2^N$ conjugate states $<k_1 \ldots k_N|$. This
gives $2^N$--dimensional Fock representation for $\t$ and $\d$
as well as for any polynomial of them.

Thus the above relations guarantee necessary physical properties
of the variables and derivatives (viewed as creation and annihilation
operators). But, is the reverse statement true? In other words, is the
above Grassmann algebra uniquely defined by some natural physical
requirements? Speaking rigorously, we should use the term `Grassmann
algebra' only for the algebra of the variables $\t$ while the algebra
of both variables and derivatives is better to be called `Grassmann
differential calculus'. For the sake of brevity we will often neglect
this tiny distinction.

It is easy to prove that the algebra of the variables $\t_i$ is
uniquely defined by the condition that
\be
\lb{w2}
(\sum_{i=1}^{N} c_{i}\t_{i})^{2}=0 \; .
\ee
The same is true for the derivatives $\d_i$. This condition is
quite natural as it says that any superposition of the Fermi
states also satisfies the exclusion principle. More difficult is
to understand the nature of Eq.(\ref{w13}). One may try a more
general but still natural and simple bilinear {\em ansatz}
$$\d_{i}\t_{j}=\delta_{ij} + R_{ij}^{kl}\t_{l}\d_{k} \; ,$$
where the summation over $k, l$ is assumed. However it is not
difficult to prove that Eqs.(\ref{w1}, \ref{w11}) and associativity
of the multiplication in our algebra require
$$R_{ij}^{kl} = - \delta_{ik} \delta_{jl} \;.$$
Thus Eq.(\ref{w13}) is the most general bilinear expression
for $\d_i \t_j$ with the anticommuting Grassmann variables and
derivatives. One might try to further generalize this result
but we end our discussion of the Grassmann case and turn to
more fundamental, paragrassmann generalizations. Our approach to
constructing paragrassmann algebras and calculus is to follow,
as close as possible, to the Grassmann pattern briefly described
above (for a more mathematical treatment and physics applications
of the Grassmann calculus see Ref.\cite{ber}).

Let us first fix the notation.
An algebra generated by $N$ $p$-nilpotent variables $\t_i$,
with some (to be found below) generalized `commutation' relations
between them, will be denoted by $\GG$.
An algebra generated by both $\t_i$ and $\d_i$
will be denoted by $\PP$.
Both $p$ and $N$ could be omitted in clear situations.
So the fermionic algebra given by Eqs.(\ref{w1}, \ref{w11}, \ref{w2})
is called $\Pi_1(N)$ in our notation,
while the algebra generated by $\t$'s only, i.e. the standard
Grassmann algebra known in the last century, is denoted by $\G_1(N)$.
We usually call `paragrassmann algebra' both $\G$ and $\P$
though the latter is in fact a generalization of the Grassmann calculus
and should be called paragrassmann calculus.

It is clear that $\G_p(1)$, the algebra of one variable $\t$, is a well
known object $k[\t]/(\t^{p+1}$).
Here $k$ is the principal field (or ring), say the field of
complex numbers or complex functions. The latter is needed for
constructing a para-super geometry but we will not address this extremely
difficult topic here; first steps in this direction were attempted in
\cite{fik2}. The algebras $\P_p(1)$ were described in \cite{fik1}.
As was shown there, each algebra $\P_p(1)$ is defined by $p$ independent
complex numbers $\a_n$, so that (recall that $\d_i |0> = 0$):
$$\d \t^n|0> = \a_n \t^{n-1}|0>, \;\; n = 1, \ldots , p \;.$$
All algebras with nonvanishing $\a_n$ are called nondegenerate, and
they are mathematically equivalent. So we call them `versions' of
the same algebra This does not necessarily imply
their physical equivalence. For example, for the physicist it
would be strange to identify parabosons with parafermions
(see \cite{green}, \cite{volk}, \cite{kam}). Nevertheless, the algebras
describing them may be obtained by our general procedure and they
correspond to two different non-degenerate systems of the parameters
$\a$.

An alternative and equivalent way of defining the para-derivative
$\d$ is to write the operator (matrix) relation
$$\d \t = b_0 + b_1 \t \d + \cdots + b_p \t^p \d^p\;, $$
where the parameters $b_i$ satisfy one equation, which follows from
the identity $\d \t^{p+1} \equiv 0$. In terms of this relation
the simplest version is defined by the bilinearity requirement
which by simple calculations and normalizing $\t$ reduces to
         $$\d \t = 1 + q \t \d \;,$$
where $q$ is any primitive root of unity (i.e. $q^{p+1}=1$
and $q^k \neq 1$ for smaller $k$). Thus the bilinear version
of the paragrassmann algebra $\P_p(1)$ is the closest one to
the Grassmann algebra $\P_1(1)$.

Here we will discuss the algebras $\GG$. As we have shown in the
above references (and will be demonstrated more generally below)
there are many nonequivalent algebras for given $p$ and $N$ but there
exist important exceptions to this general rule. This statement is
true even for the simplest bilinear algebras that we have studied
so far. The non-bilinear algebras are not well understood but we
hope that the bilinear algebras are most important for physics
applications. They are also much simpler to work with and natural
physical (or geometrical) conditions usually lead to the bilinear
case (see \cite{fik2}).

The first quite natural requirement for paragrassmann
algebras of many variables is that
any linear combination of $\t_{i}$ should be $p$-nilpotent,
which means that  for any choice of the complex numbers $c_i$
\be
\lb{w21}
(\sum_{i=1}^{N} c_{i}\t_{i})^{p+1}=0 \; ,
\ee
and thus the variables $\ti$ generate the linear space over complex
numbers. As the coefficients $c_i$ are simply commuting,
this condition is equivalent to the following set of
$(p+1)$-linear equations
$$
\sum_{\s} \t_{\s(i_0)}\dots \t_{\s(i_p)}=0 \;,
$$
where the sum is taken over all different permutations, $\s$, of the set
$$I=\{i_0,\dots,i_p \}\;,\;i_k\in \{1,\dots,N \}.$$
One could try to treat these monstrous sums as analogues of the
anticommutators but that would fast lead to a dead end.
So, to proceed with developing a paragrassmann calculus some further
restrictions must be imposed on the nilpotent variables.
Here is a branching point since these restrictions are to be made by
hands and depend on what do we want of the paragrassmann algebras.
To make our choice more clear we remind some history of the paragrassmann
algebras and motives of our investigation.

In fact, the history of the paragrassmann algebras can be traced back to
the earliest days of quantum theory. A sort of a para-Clifford algebra
(with unipotent variables) was introduced by Hermann Weyl in twenties.
In sixties, Julian Schwinger used Weyl's ideas for regularizing quantum
field theories but then left the general approach and concentrated on
the Grassmann algebra.  A different development started in fifties
with the aim to clarify the parastatistics problem.
First remarkable results had been obtained by H.S.Green \cite{green}
and D.V.Volkov \cite{volk}.
The work in subsequent decades had been summarized in \cite{kam}.

The central characters of that `old testament' were  quantum fields,
subjected to parastatistics, i.e. having $p$-fold degeneracy in
symmetric (`parafermions') or skew symmetric (`parabosons') states.
Paragrassmann algebras appeared there as a somewhat auxiliary tool
in a framework of the so called {\em Green ansatz}. It consists of
representing each $\ti$ by the sum of $p$ Grassmann numbers,
$$
\ti =\sum_{\a=1}^{p} \zeta^{(\a)}_{i}\;,
$$
with the following commutation relations for the components
$$
\zeta^{(\a)}_{i}\zeta^{(\b)}_{j} =
(-)^{\delta_{\a \b}} \zeta^{(\b)}_{j}\zeta^{(\a)}_{i}\;.
$$
The $p$-nilpotency of $\t$'s is obviously guaranteed, and also it is
easy to check that
$$
[\ti , \tj] = 2\sum_{\a}
\zeta^{(\a)}_{i}\zeta^{(\a)}_{j}\;,
$$
and therefore
$$
[[\ti,\tj],\t_k]=0\;.
$$
This trilinear condition had been taken as a basic identity for
the generators, in addition to nilpotency.
Being combined with (\ref{w21}) it leads to some simplifications,
like $\ti\tj\t_k + \t_k \ti\tj =0$ for $p=2$ , but clearly can never
produce any bilinear identities and, therefore, commutation formulas.
The correspondent differential calculi are also essentially multilinear
and rather messy, making it hard to get any far-reaching results,
except for the simplest case $p=2$ .

A new philosophy was suggested by two-dimensional conformal field
theories, where fields of fractional spin appear quite naturally.
In fact, our approach (see \cite{fik}, \cite{fik1}, \cite{fik2})  arose
in the course of search for a geometric meaning of conformal algebras and
of their representations by generalized differential operators.
A central object in this context is a fractional derivative
${\cal{D}},\;({\cal{D}}^{p+1}=\d_z) $, that transforms fields of the
conformal weight $\lambda$ into ones of the weight $\lambda+1/(p+1)$ .
Such a derivative has a natural paragrassmann representation
\be
\lb{der}
{\cal{D}}=\d + \kappa \t^{p} \d_z\;,
\ee
where $\d$ and $\t$ generate what we call $\Pi_p(1)$, and $\kappa$
is a normalization coefficient.
(Note that this way of writing a formal root of any operator was more
or less widely known for some time; it is certainly not our invention.)
It is seen from (\ref{der}) that $\d$ and $\t$ must have conformal
dimension $ \pm 1/(p+1)$.  As the Grassmann numbers have dimension 1/2,
Green's construction is incompatible with the above consideration.
Once we insisted on importance of the fractional derivative we had to
try a new start in the paragrassmann business, throwing away the Green
ansatz however clever it was.

Note that some similar attempts can be found in recent literature, usually
in the context of so called `fractional supersymmetry' (references can be
found in our papers quoted above). By the way, we think that the terms
`fractional supersymmetry', and even more, `fractional Grassmann algebras'
sound a bit strange because, if anything
there is fractional, that is the conformal dimension or the derivative
but the symmetry and algebras are no more `fractional' than the standard
supersymmetry which is never called `fractional'.
For this reason, we prefer to speak about {\it para}supersymmetry and
{\it para}grassmann algebras, keeping in mind, that they have nothing to
do with Green's approach (with these reservations, we also sometimes use
the term `fractional' for paragrassmann algebras and symmetries).
 Though the simplest things, like Eq.(\ref{der}) or particular
paragrassmann algebras, were known for some time, it seems that the full
depth and richness of the subject has not been realized so far.
This is particularly true for the paragrassmann algebras with many
variables which are the main subject of this report.

Another desirable property of the paragrassmann algebras dictated by
applications is existence of bilinear identities between the variables
and, even more important, the possibility of a normal ordering of them.
To formulate this more precisely let us denote by $\G^{(m)}$ the subspace
of $m$-linear polynomials in $\ti$. Then the set of ordered monomials
\be
\lb{nord}
\t_{i_1} \dots \t_{i_m}  \;,\;i_1 \ge i_2 \ge \dots \ge i_m
\ee
must form an additive basis of $\G^{(m)}$. A sufficient condition for
this is provided by `commutation' relations of the type ($i<j$):
\be
\lb{rcom}
\ti \tj = \sum_{k<l\leq j} R^{kl}_{ij} \t_l \t_k \;,\;l>k\;.
\ee
The restrictions on the indices allow one to reorder
the variables $\ti$ even for infinite number of the variables.
More general bilinear relations might be used with this aim but
we need not consider them here.
The matrix $R$ must satisfy some additional relations similar to those
known in the theory of the Yang--Baxter equations and of the quantum
groups (see \cite{Ludwig}, \cite{Luis}, \cite{Ivan}).
We will not discuss these topics here but it is worth noticing that the
commutation relation of this sort are usually discussed without
adding the nilpotency condition (\ref{w21}) which is in fact the heart
of our approach.

Note  that the requirement of normal ordering guarantees
a `para-supersymmetric' structure of the space $\GG$ generalizing the
supersymmetric structure of the space $\G_1(N)$ in the following sense.
Let us divide the space of all polynomials
of $\t_i$ into subspaces having the same degree modulo $p+1$:
$$
\G^{[M]} = \bigoplus_{m=M\;mod(p+1)} \G^{(m)}\;,\;\;
\g_m = {\rm dim } \G^{(m)}\;.
$$
Then the dimensions of all $\G^{[M]}$ do not depend on $M$ and
are equal to $(p+1)^{N-1}$. This immediately follows from the identity
$$
(1+x+ \dots +x^p )^N = \sum_{m \ge 0} \g_m x^m\;.
$$
This property of the Grassmann algebras explains their usefulness for
formulating supersymmetries, we suggest to call it `rudimentary
supersymmetry' (meaning that it needs no Lagrangians or Hamiltonians
and is formulated in terms of the space of states). The generalization
suggested here might be called a `rudimentary para-supersymmetry' and
it probably will constitute a basis for a more general approach to
para-supersymmetric physical systems introduced earlier in terms of
special representations for $\t$-variables (see \cite{rub} and
further references in our papers quoted above).

Now we have to recall that our main aim is to satisfy the nilpotency
condition (\ref{w21}). However, this problem proves to be very difficult
to solve for the general bilinear commutation relations (\ref{rcom}).
To be able to find algebras for arbitrary values of $p$ and of $N$
we thus are forced to make a simple enough {\em ansatz} for $R$.
Namely, we suppose that there exists a basis $\t_1, \dots , \t_N$
in $\G^{(1)}$ for which the $R$-matrix is diagonal
\be
\lb{diacom}
R^{kl}_{ij} = \delta^k_i \delta^l_j \;r_{ij}\;,
\;\; {\rm i.e.}\;\; \ti \tj =r_{ij} \tj \ti \;.
\ee
In other words, we impose a multi-grading
${\rm deg}_i \tj =\delta_{ij}$ (and, correspondingly,
${\rm deg}_i \d_j =\delta_{ij}$  for $\Pi $),
which must be preserved under commutation.
It seems to be the strongest requirement we can impose.

In terms of the commutation relations (\ref{diacom}) our problem
can now be formulated in the following form:
{\it to find all admissible sets of $r_{ij}$}, i.e.
those for which all the equations (\ref{w21}) are satisfied.

To formulate this problem in more precise terms let us define
$$
\{\t_1^{(k_1)},\dots,\t_N^{(k_N)} \} =
\sum_{\s} \t_{\s(1)}\dots \t_{\s(p+1)}\;,
$$
where the sum is taken over all permutations of the multiset
$K=\{1^{k_1},\dots,N^{k_N} \}\;\;, \\
k_i\ge 0\;\;,\;k_1+\dots+k_N=p+1\;$ (the meaning of the notation
for the multiset or the multi-index must be clear -- each number
$i$ is taken $k_i$ times.
These brackets naturally appear in calculating
$$
(\t_1 + \dots +\t_N)^{p+1} =
\sum_{K}
\{\t_1^{(k_1)},\dots,\t_N^{(k_N)} \}\;\;.
$$
Here the sum is taken over all $(p+1)$-submultisets $K$ (introduced above)
of the big multiset $Z=\{1^p, \dots, N^p\} $.
Using the commutation relations we can write all these brackets in
the normal ordered form:
$$
\{\t_1^{(k_1)},\dots,\t_N^{(k_N)} \} =
P_K (r_{ij})\;\t_N^{k_n} \dots \t_1^{k_1}\;.
$$
where $P_K$ are polynomials of $r_{ij}$. To guarantee the nilpotency
property $r_{ij}$ have to be chosen so that all polynomials are zero.
At the moment, we do not know explicit expressions or simple recurrence
formulae for these polynomials and, of course, we can not make any
general statement about their zeroes.
So we will try to move step by step attempting
to reduce the general case to the simplest ones.

The simplest thing is to consider a subalgebra $\G(2)$
generated by two variables.
It is completely solved by the following

\bf Lemma 1. \rm $r_{ij}$ is a primitive root of unity, i.e. has the form
\be
\lb{rq}
r_{ij} = q^{\rij}\;\;,\;\; q=exp(\frac{2\pi i}{p+1}) \;,
\ee
where $\rij \in {\bf I}_{p+1}$ -- the multiplicative group of
invertible elements of the ring ${\bf Z}_{p+1}$. In other words,
the set of numbers which are less than $p+1$ and are relatively prime
to it (i.e. having no common divisors with $p+1$).
The statement of the lemma follows directly from the simple fact:
\be
\lb{p1}
P_{k,\; p+1-k} (r) = \frac{(p+1)_r!}{(p+1-k)_r !\;(k)_r !}\;,
\ee
where we use the standard notation
$$
(n)_r = \frac{r^n - 1}{r-1}\;,\;\;(n)_r !=(n)_r (n-1)_r \dots (1)_r \;.
$$
Polynomials (\ref{p1}) are the well-known Gauss polynomials (or
the $r$-binomial coefficients), and clearly all of them are zero iff
$r^{p+1}=1$ and $r^k \not= 1\;, \; k<p+1$.

Some words should be said about the groups ${\bf I}_{p+1}$
because they play a crucial (though not obvious) role in the whole
our construction.
They are abelian groups of order $\varphi (p+1)$ (this is the Euler
function giving the number of positive integers that are less than
$p+1$ and relatively prime to it) and therefore must be isomorphic
to direct products of simple cyclic groups ${\bf C}_{p^n}$.
As they are not well known to physicists, we present below a few
examples of them (with the correspondent residues in parentheses):
$$
\begin{array}{lll}
{\bf I}_3 & \simeq & {\bf I}_4\;\simeq \;{\bf I}_6 \; \simeq \;
 {\bf C}_2 (1,-1)\;,  \\
{\bf I}_5 & \simeq & {\bf C}_4 (1,2,-1,-2)\;,\\
{\bf I}_7 & \simeq & {\bf C}_2 (1,-1) \times {\bf C}_3 (1,2,4)\;
\simeq \; {\bf C}_6\;,\\
{\bf I}_8 & \simeq & {\bf C}_2 (1,-1) \times {\bf C}_2 (1,3)\;,\\
{\bf I}_9 & \simeq & {\bf C}_2 (1,-1) \times {\bf C}_3 (1,-2,4)\;
\simeq \; {\bf C}_6\;,\\
{\bf I}_{10} & \simeq & {\bf C}_4 (1,3,-1,-3)\;,\\
{\bf I}_{11} & \simeq & {\bf C}_2 (1,-1) \times {\bf C}_5 (1,-2,4,-8,5)\;
\simeq \; {\bf C}_{10}\;,\\
{\bf I}_{12} & \simeq & {\bf C}_2 (1,-1) \times {\bf C}_2 (1,5)\;,\\
{\bf I}_{13} & \simeq & {\bf C}_3 (1,3,9)\times {\bf C}_4 (1,5,-1,-5)\;
\simeq {\bf C}_{12}\;\;,\\
{\bf I}_{14} & \simeq & {\bf C}_2 (1,-1) \times {\bf C}_3 (1,-2,4)\;
\simeq \; {\bf C}_6\;,\\
{\bf I}_{15} & \simeq & {\bf C}_2 (1,-1) \times {\bf C}_4 (1,2,4,8)\;,\\
{\bf I}_{16} & \simeq & {\bf C}_2 (1,-1) \times {\bf C}_4 (1,5,9,13)\;,\\
{\bf I}_{17} & \simeq & {\bf C}_{16}\;,\\
{\bf I}_{18} & \simeq & {\bf C}_2(1,-1) \times {\bf C}_3 (1,7,13)\;,\\
{\bf I}_{19} & \simeq & {\bf C}_2 (1,-1) \times {\bf C}_9\; \simeq \;
{\bf C}_{18}\;,\\
{\bf I}_{20} & \simeq & {\bf C}_2(1,-1) \times {\bf C}_4 (1,3,9,7)\;,\\
\dots &  &  \dots \dots \dots \\
{\bf I}_{24} & \simeq & {\bf C}_2 (1,-1) \times {\bf C}_2 (1,5) \times
{\bf C}_2 (1,7)\;.
\end{array}
$$

Thus, all $\G_p (2)$ are very simply classified.
Once it has been done, a possibility of an inductive construction
of $\GG$ arises (special examples of this new general construction
were given in our papers \cite{fik}, \cite{fik1}; the reader is advised
to consult these papers for a better understanding of the general idea).
Thus we start with a single $\t$,
then replace it by a linear combination of two $\t$'s,
generating $\G(2)$, then this operation can be applied to any of these
new $\t$'s, and so on (we will call this a {\em telescopic construction}).
In this way, after $N-1$ steps we get a paragrassmann algebra with $N$
variables. This algebra can be visualized by a tree having
the root, $N$ free ends numbered from left to right by
$1, 2 \ldots N$, and $N-1$  vertices, labelled by some integers
$\r_m \in {\bf I}_{p+1}$. We assume that the branches growing from the
same vertex are ordered from {\it left} to {\it right}).
The commutation relations can be read from the tree as follows:
if the end $i$ belongs to a {\it left} sequence of branches and
the end $j$ belongs to a {\it right} sequence of branches both growing
from the same vertex labelled by $\r_m$, then
$$
\ti \tj = q^{\r_m} \tj \ti \;.
$$
Let us call paragrassmann algebras of the described kind {\it maximal},
in a sense to be clarified later.

Clearly, the trees that can be transformed into each other by a sequence
of transformations like
``$\r_m \rightarrow -\r_m$ ; $left \leftrightarrow right$ at
the vertex $m$'' correspond to equivalent algebras.
The algebras obtained by a re-numbering $\t$'s are also identical.
It is also reasonable to factor out the action of the group
${\bf I}_{p+1} \;: \rij \rightarrow \s\rij\;,\;\s\in{\bf I}_{p+1}$
(because $q$ may denote any primitive root of unity in (\ref{rq})).
 Note, that $\s =-1$ is equivalent to the re-ordering, so in fact we
have factored out the group
$S_N \times {\bf I}_{p+1}/{\bf C}_{2}$.
The problem of equivalence becomes more and more complex as $p$
and $N$ grow but it is clear that there {\it exist}
non-equivalent algebras for any values of $p$ and $N$.
This is a new feature of the paragrassmann algebras as compared to
the Grassmann case, where the algebra was completely determined by
$N$ ($p=1$). We will show that the paragrassmann algebras are completely
fixed by $N$ only for $p=2,3,5$ when the unique equivalence class exists.

Remark in passing that the integer-valued matrix $\rij$ has a deeper
meaning  than  $r_{ij}$ defined in (\ref{diacom}). If we denote
$w^a =\t_1^{a_1}\t_2^{a_2}\dots \t_N^{a_N}$ and
$w^a w^b =q^{[a,b]} w^b w^a$ , then it is easy to show that
$$
[a,b] = \rij a_i b_j\;.
$$
So $\rij$ plays the role of a skew-symmetric 2-form in the vector space
of vectors $a$ enumerating the monomials in $\GG$.

Let us return to the above telescopic construction.
It is clear that it can be generalized in the following way.
Namely, if we already have got two paragrassmann algebras
$\G_p(N)$ and $\G_p(M)$, then another algebra
of $N+M-1$ variables can be obtained via replacing some $\ti$
of the first algebra by an arbitrary linear combination of $M$ generators
of the second one. Let us call this new algebra the {\it telescoping
product} of the two algebras. The best way to describe it is to use
tree graphs similar to that described above.
We will not go into a detailed description of this construction.
The only important thing is to understand that, by the telescoping
procedure, all possible algebras can be made of irreducible,
indecomposable blocks having only one vertex.
Such building blocks we call {\it minimal} algebras. In summary,
the structure of the complex paragrassmann algebras depends not only
on their minimal building blocks but also on the way of combining
them into the structure represented by the tree diagram described above.

Thus, to classify paragrassmann algebras we have to find all minimal
algebras and to find criteria of equivalence of algebras corresponding
to different trees. We will not go into a discussion of the equivalence
problem and only briefly summarize what we know about minimal algebras.
A series of the minimal algebras for any $p$ is given by the described
above algebras $\G_p(2)$.
Until recently we believed that there is no other minimal algebras.
However, below we present an example of a minimal algebra $\G_4(4)$.
In fact, it is the only example we presently know.
We think that algebras of this sort are really exceptional and do not
exist for large values of $p$ and $N$.

Let us proceed with analyzing $N=3$ subalgebras.
We believe that there are no minimal algebras with three
generators (in fact, we have an incomplete proof of this statement
which is valid for almost all values of $p+1$). This means that, for
any admissible set of $r_{ij}$, two of the three numbers
$r_{ij},r_{jk},r_{ki}$ must be mutually inverse for any choice of
$i, j, k$ (mind that the order of the indices is cyclic).
If we visualize the paragrassmann algebra with $N$ variables by
a complete (full) graph
with $N$ vertices and $N(N-1)/2$ oriented edges $i \rightarrow j$
labelled by the numbers $\rij$, the above statement means that
all triangles in the graph are isosceles (e.g. $\rij = \rik$).

We will not present here even a sketch of the proof and only give
the first step of it which is useful by itself. This is

{\bf Lemma 2}. $\;\;$
The equality
\be
\lb{lem2}
r_{ij}=r_{jk}=r_{ki}=q \; ,
\ee
where $q$ is any primitive root of unity (i.e. $q^{p+1}=1$ and
$q^k \neq 1$ for smaller $k$) {\em is not admissible}.

This can be seen from the following expression for the
simplest $N=3$ polynomial
\be
\lb{p11}
P_{p-1,1,1}(q,q,q^{-1}) = (p+1)(q+1) \neq 0\;,
\ee
where the polynomial is defined by
$$\{\t_i^{(p-1)}, \t_j, \t_k\} = P_{p-1,1,1}(r_{ij},r_{jk},r_{ik})
\ti^{(p-1)} \tj \tk$$
and can be calculated recursively.

An immediate consequence of this Lemma is the complete classification for
the cases $p=2,3,5$. Indeed, in these cases $\rij = \pm 1$, and so
precisely two of the three numbers in (\ref{lem2}) are equal.
Let us assume that they are $r_{ij}$ and $r_{jk}$.
If so, define the ordering of the vertices in the graph  by saying that
$i\prec j\prec k$.
It is easy to show that the transitivity of the ordering follows from
the Lemma and so all $\ti$ are ordered. Then we obviously can re-number
the $\t$'s so that the $\prec$-relation matches the standard
lexicographical ordering of the indices.
Therefore for $p=2,3,5$ paragrassmann algebras $\GG$
are uniquely defined for any $N$ by the simple $q$-commutation relations
$$
\ti \tj = q \tj \ti \;,\;i<j\;.
$$
Of course, these relations define paragrassmann algebras for any
values of $N$ and $p$ but for $p\neq 2,3,5$ they are not unique
as we have shown above.

Moreover, there exists at least one minimal algebra not coinciding
with $\G_p(2)$. Consider possible choices of the $\r$-matrices
for $N=4$, $p=4$
(satisfying all the restrictions formulated above, including
the requirement that all triangles are isosceles). Then we have
\be
\lb{nz}
\r_{12}=\r_{14}=\r_{34} =1 \;\;,\;\;
\r_{13}=\r_{23}=\r_{24} = a \;,
\ee
where $a=1,2,3$.
The choice $a=1$ gives the above non-minimal algebra and it is
easy to prove that for $a=3$ the matrix is not admissible.
But a direct check shows that {\it for $a=2$ the matrix
{\rm (\ref{nz})} is admissible, and therefore
the algebra described by it is a minimal paragrassmann algebra.}

As it is easy to see, this algebra is {\it exceptional} in any possible
sense. Indeed, the case $p=4$ is exceptional, since in it there is only
\it one \rm
(modulo cyclic permutation) polynomial with four non-zero $k$'s, namely
$P_{1,1,1,2}$, and it prohibits only one of the two possibilities ($a=3$).
As the number of non-equivalent polynomials is fast growing with $p$,
it seems unlikely that something similar could happen for large $p$,
and minimal algebras must really be very rare exceptions. Unfortunately,
at the moment we do not know how to treat the general case.

Summarizing present status of the paragrassmann algebras,
we wish to emphasize the fact that there exist some algebras for
any $N$ that are almost as simple as the Grassmann algebras but, in
general, the `para-world' is much richer than the `super-world'.
Two main unsolved problems of the $\GG$ algebras theory are:
1.~to construct {\em all} possible minimal algebras;
2.~to find to what extent the diagonalizable $R$-matrices represent
the class of all admissible ones.

The next topic that has to be discussed is how to construct differential
calculi like those suggested in \cite{wz}. This problem has been
essentially solved in our papers \cite{fik}, \cite{fik1}. There it
was demonstrated that a larger set of the algebras $\P$ exists in our
case, and we succeeded in classifying them by using
a species of tree graphs (see \cite{fik1}, the general consideration
of this reference can easily be adapted to general algebras $\G$
constructed here). Earlier we have discussed in some detail the
relation of the paragrassmann algebras to quantum groups and
$q$-deformed oscillators with $q = {\rm root\; of\; unity}$
(see \cite{Ludwig}, \cite{Luis}, \cite{Ivan}, \cite{Alan}).
We have shown that bilinear paragrassmann algebras are most
directly related to the representations of quantum groups with
$q = {\rm root\; of\; unity}$. The new results on $\GG$ presented above
give even more support to this conclusion. The connection is not so
clear for non-bilinear algebras (not treated here) and, in addition,
it is not so direct as it seems at first sight (because of
our emphasis on the nilpotency condition (\ref{w21})). Due to this,
our algebras in simplest cases are indeed direct generalizations
of the Grassmann calculus (see also \cite{fik2} for a definition of
integration in paragrassmann variables and of simplest paraconformal
transformations -- para-translations, para-inversions, etc.).

Nevertheless, possible applications of the paragrassmann algebras
may be in the problems in which quantum groups are useful. These
include rational conformal theories \cite{Ludwig}, \cite{Luis},
\cite{Ivan} and, more generally, integrable models \cite{Ludwig},
\cite{Andre}. We hope that the extension of possible dynamical
symmetries (para-supersymmetries, para-conformal symmetries, etc.)
which is provided by paragrassmann calculus may prove
important for applications to physical systems even in more than
two space-time dimensions.

The final version of this report is significantly different from
that presented at the Smorodinsky Workshop. We have omitted almost
all published results and added new results obtained after
the Workshop.

For useful discussions of the preliminary versions of our report
we would like to thank A.P.Isaev and A.A.Vladimirov. This work
would be never finished without continuous kind support of V.de Alfaro
and A.Salam to whom one of the authors (A.T.F.) expresses his deepest
gratitude. Useful discussions of the results with L.Alvarez-Gaume,
M.Mintchev and I.Todorov are acknowledged. Kind hospitality and
financial support of ICTP (Trieste), the Turin Section of INFN, and
of the Physics Department of the University of Turin are appreciated
by one of the authors (A.T.F.).


\end{document}